# Effect of interfacial $Fe_3O_4$ nanoparticles on the microstructure and mechanical properties of textured alumina densified by ultrafast high-temperature sintering

Rohit Pratyush Behera[1], Andrew Yun Ru Ng[2], Zehui Du[2,3], Chee Lip Gan[2,3], Hortense Le Ferrand[1,3*]

[1] *School of Mechanical and Aerospace Engineering, Nanyang Technological University, 50 Nanyang avenue, Singapore 639798*
[2] *Temasek Laboratories, Nanyang Technological University, 50 Nanyang Drive, Singapore 637553*
[3] *School of Materials Science and Engineering, Nanyang Technological University, 50 Nanyang avenue, Singapore 639798*

* Corresponding author: hortense@ntu.edu.sg

## Abstract

Alumina microplatelets coated with a small amount of $Fe_3O_4$ can be oriented via a rotating magnetic field to create texture. After ultrafast high-temperature sintering (UHS), Fe atoms are found at the grain boundaries and within the grains, influencing the mechanical properties. Here, we compare the microstructure and mechanical properties of textured alumina prepared with and without $Fe_3O_4$ and sintered using UHS or conventional sintering (CS). Microstructural analysis using electron backscattering diffraction (EBSD) indicates that $Fe_3O_4$ induces crystallographic defects in the ceramic after UHS. Nanoindentation measurements enlighten that the presence of $Fe_3O_4$ leads to plastic flow that increases the energy dissipation, reaching ~122 % at a maximum load of 1900 mN compared to pristine samples. Overall, due to the concentrated effects of $Fe_3O_4$ after UHS, the flexural strength and fracture toughness values are higher than the other two samples, reaching values of ~287 MPa and 7 MPa.m$^{0.5}$, respectively. These results could be leveraged to produce stronger and tougher ceramics.

Fast sintering processes, such as spark plasma sintering [1], flash sintering [2], and ultrafast high-temperature sintering (UHS) [3] etc., typically employ a heating rate varying from 100 to 20,000 °C/$min$ and short isothermal sintering times depending on the process, to sinter a variety of ceramics. A typical advantage of such fast processes is to suppress grain growth and promote densification that is often aided by pressure during sintering. For textured ceramics that have metal or metal oxide coatings, the SPS assisted by pressure has been used and is listed as a key process to retain high aspect ratio anisotropic grains, which also promotes the desired reduction of metal oxides in an inert sintering atmosphere [4–7]. As a result, materials produced by this method exhibit a rising R-curve in fracture toughness tests, thus preventing catastrophic failure along with a strength higher or similar to uncoated pristine ceramics [4–7]. Therefore, more attention has been focused on the interface between grains or grain boundaries in a polycrystalline ceramic system. However, it remains unclear if such coatings or interfacial materials could cause significant changes within the microstructure in the form of crystallographic defects when sintered at ultrafast heating rates in the absence of pressure that could also affect the mechanical properties.

Recently, pressureless UHS has been used to sinter textured alumina that uses radiation and Joule heating to sinter bulk ceramics at ultrafast heating rates with no direct passing of current through samples [8], thus eliminating the possible introduction of crystallographic defects due to external fields as in other fast sintering processes [1,2]. These textured aluminas are prepared using green bodies made of alumina microplatelets decorated with a small amount of $Fe_3O_4$ nanoparticles and oriented horizontally using magnetically assisted slip casting (MASC) in a bimodal system consisting of alumina nanoparticles dispersed homogeneously and densified via templated grain growth (TGG) [8]. The textured ceramics are remarkable because they achieve a flexural strength ~287 MPa and ~7.0 ± 0.3 MPa.m$^{0.5}$ with rising R-curve reaching crack growth toughness ($K_{Jc}$) ~7.0 ± 0.3 MPa.m$^{0.5}$ [9]. Although these properties arise in part from the texturing, the values obtained are higher than those reported in the literature for similar textured alumina prepared and sintered using other methods, such as a crack initiation toughness ($K_{Ic}$) of ~3.6±0.5 MPa.m$^{0.5}$ using a lithographic-based process and pressureless SPS [10] and micro-scale $K_{Jc}$ of ~3.3±0.2 MPa.m$^{0.5}$ using tape-casting and pressure-assisted sintering [11]. The hypothetical reason that may explain the increased mechanical properties of the textured alumina prepared using MASC and UHS is the presence of $Fe_3O_4$ nanoparticles in the green body. Indeed, it was found in the sintered ceramics that Fe atoms localised around the grain boundaries and contributed to the anisotropic microstructure and texturing [8]. However, the mechanisms by which these Fe atoms may further affect the microstructure and mechanical properties are still unknown. These could provide interesting microstructural insights for similarly coated ceramics densified at ultrafast heating rates.

Previous works using UHS have reported the reduction of vaporisation or loss of low-temperature interfacial materials, such as Li-based garnet in battery materials [3] and Ni-based superalloy in $Al_2O_3$ cermet [12]. Moreover, UHS has been claimed to be an out-of-equilibrium process where microstructural defects have been recognised around grain boundaries due to ultrafast heating rates [13,14]. However, no role of interfacial material in creating microstructural crystallographic defects has been studied yet in UHS to the best of the authors' knowledge. Defects are omnipresent in the microstructure of synthetic materials since most processes are far from equilibrium and take place at temperatures > 0 K. In the past years, there has been renewed interest in introducing crystallographic defects in ceramics in order to tune their mechanical properties, especially to induce plasticity and increase their fracture toughness [15–18]. These defects can be high-angle grain boundaries (HAGB), and low-angle grain boundaries (LAGB), stacking faults, twin boundaries, line defects (dislocations), etc. However, to achieve even a tiny plastic effect, a typical defect or dislocation density of $\geq \sim 10^{13}/m^2$ is required, in contrast to the general $\sim 10^9$ to $10^{10}/m^2$ present after high-temperature sintering [19]. Therefore, several strategies and experimental protocols have been introduced to enrich ceramics with dislocations for tailoring functional and mechanical properties, such as light illumination in ZnO single crystals [20], shot blasting and high-temperature annealing in $Al_2O_3$ single crystals [21]. Hence, studying the presence and quantifying of crystallographic defects in the textured alumina prepared using MASC and UHS could help understand the underlying mechanisms behind the unusually high mechanical properties and leverage future strategies to enhance these properties further, in particular, the fracture toughness.

This work, therefore, studies the effects of Fe atoms on the microstructure and mechanical properties of textured alumina after UHS. Three types of textured alumina samples were prepared and compared: (i) with $Fe_3O_4$ and sintered using UHS (UHS, with $Fe_3O_4$), (ii) without $Fe_3O_4$ and sintered using UHS (UHS, without $Fe_3O_4$), and (iii) with $Fe_3O_4$ and sintered using conventional sintering CS (CS, with $Fe_3O_4$).

## Experimental

**Material fabrication**: Bulk textured alumina samples were prepared using MASC and UHS via templated grain growth (TGG) **(Fig. 1a)**. Alumina microplatelets of ~6.36 μm diameter (Kinsei Matec) were initially decorated with 0.75 vol % nanoparticles of $Fe_3O_4$ (10 nm, EMG-705, Ferrotec) to make them magnetically responsive [22]. The newly obtained magnetised microplatelets were then suspended in water with alumina nanoparticles of diameter ~100-120 nm (AERODISP® W 440, Evonik; AP-D powder, Struers) at alumina microplatelet to nanoparticle ratio of 5:95 and 55 % total solid loading. The suspension was then cast onto a porous gypsum substrate while a rotating magnetic field, $H_{rot}$ of ~50 mT was applied to orient the microplatelets horizontally following the MASC process [23]. The cast sample was then oven-dried and sintered using UHS to obtain textured alumina by TGG [24,25]. Previous work reported optimising the UHS parameters to achieve 95-98 % relative density and texture [8]. **Figure 1b** shows the typical temperature-time readings obtained using controlled power-time readings, with representative optical images of green and sintered bodies in **Figure 1c** to highlight the capability of sintering bulk samples for macroscopic mechanical testing.

**Material characterisation**: The density was measured using the Archimedes principle [25]. The microstructural features were obtained by electron microscopy (FESEM, JEOL 7600F). The grain and sub-grain features, along with quantitative values and elemental distribution, were obtained by electron backscatter diffraction (EBSD) and energy dispersive X-ray (EDX), respectively, using electron microscopy (JEOL 7800F Prime). A cube-corner nanoindenter (Triboindenter TI-950, Hysitron, Bruker) was used to investigate the energy dissipation due to fracture and to decipher plastic flow events under the indention load-contact depth curve. Finally, the fracture toughness and flexural strength were measured using a 3-point bending test setup (AG-X, Shimadzu), followed by observations of crack paths and small-scale crystallographic defects using high-resolution electron microscopy (Zeiss crossbeam 540). At least 3 samples were tested to ensure repeatability and reproducibility. Additional characterisation details and testing parameters have been described thoroughly in the **supplementary material**.

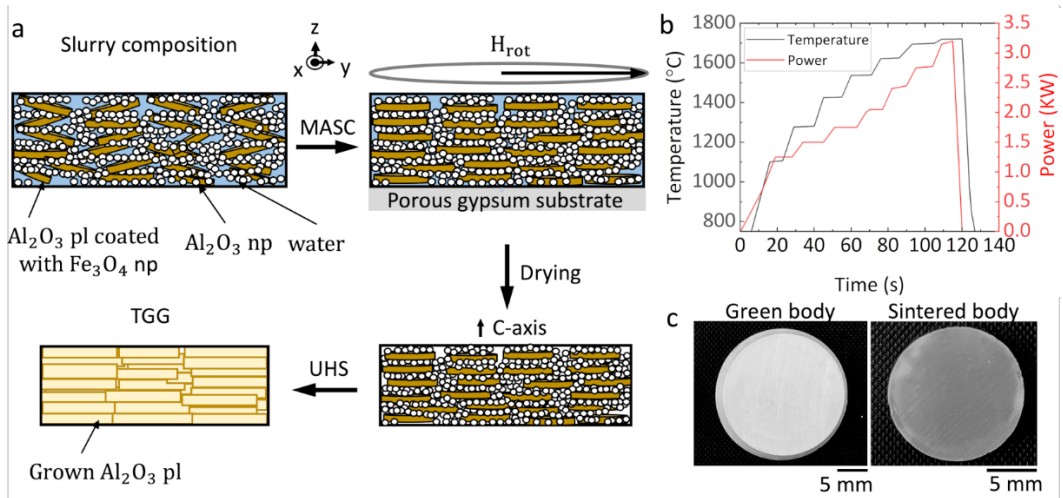

Figure 1: **Fabrication of textured alumina using MASC and UHS. (a)** Schematic illustration of MASC, where $Al_2O_3$ microplatelets (pl) coated with $Fe_3O_4$ nanoparticles (np) are suspended in water with $Al_2O_3$ np, casted on a porous gypsum under a rotating magnetic field $H_{rot}$, dried and densified via TGG. **(b)** Representative power-time readings for temperature-time profile used for UHS. **(c)** Optical images of macroscopic green and sintered bodies.

## Results and Discussion

To study the microstructural features, the textured alumina prepared by MASC and UHS were compared with samples prepared by simple casting without $Fe_3O_4$ and sintered at the same conditions using

UHS, and samples prepared by MASC and sintered using CS at 1600 ℃ for 10 h (**Fig. 2**). The relative densities of the samples were measured to be ∼96, 89 and 91 % for UHS, with $Fe_3O_4$; UHS, without $Fe_3O_4$; and CS, with $Fe_3O_4$, respectively. The low relative density in UHS, without $Fe_3O_4$, could be due to the difference in external field forces involved [26], which is only gravity in this case, in addition to the absence of $Fe_3O_4$, compared to UHS, with $Fe_3O_4$. However, no significant variance in porosity or packing fraction is observed within the microstructure of the sample (see supplementary material, **Fig. S1**) that could be present due to changes in the external field used for orientation [26]. The UHS, with $Fe_3O_4$, showed enhanced mass transport in an inert atmosphere due to the conversion of $Fe^{3+}$ to $Fe^{2+}$ discussed previously [8,27,28]. The band contrast micrographs obtained using EBSD show the presence of much higher LAGB (2-10°) ∼25.2% in the samples UHS, with $Fe_3O_4$ than for the samples UHS, without $Fe_3O_4$: ∼1.23%, and the samples CS, with $Fe_3O_4$: ∼4.74% (**Fig. 2a**). This is interesting as such dopants like $Fe_3O_4$ are known to restrict grain growth by preferring special grain boundaries like the LAGB [29]. Furthermore, the LAGBs do not show any apparent misorientations in the samples UHS, with $Fe_3O_4$ as seen from the inverse pole figure (IPF) maps, and retain a high multiple uniform density (MUD) of ∼73.75, whereas for the two other samples, the MUD is lower (**Fig. 2b,c**). Moreover, the concentration of Fe atoms in the microstructure of the UHS, with $Fe_3O_4$ samples, is higher compared to the other samples, as revealed by the elemental mapping (**Fig. 2d**).

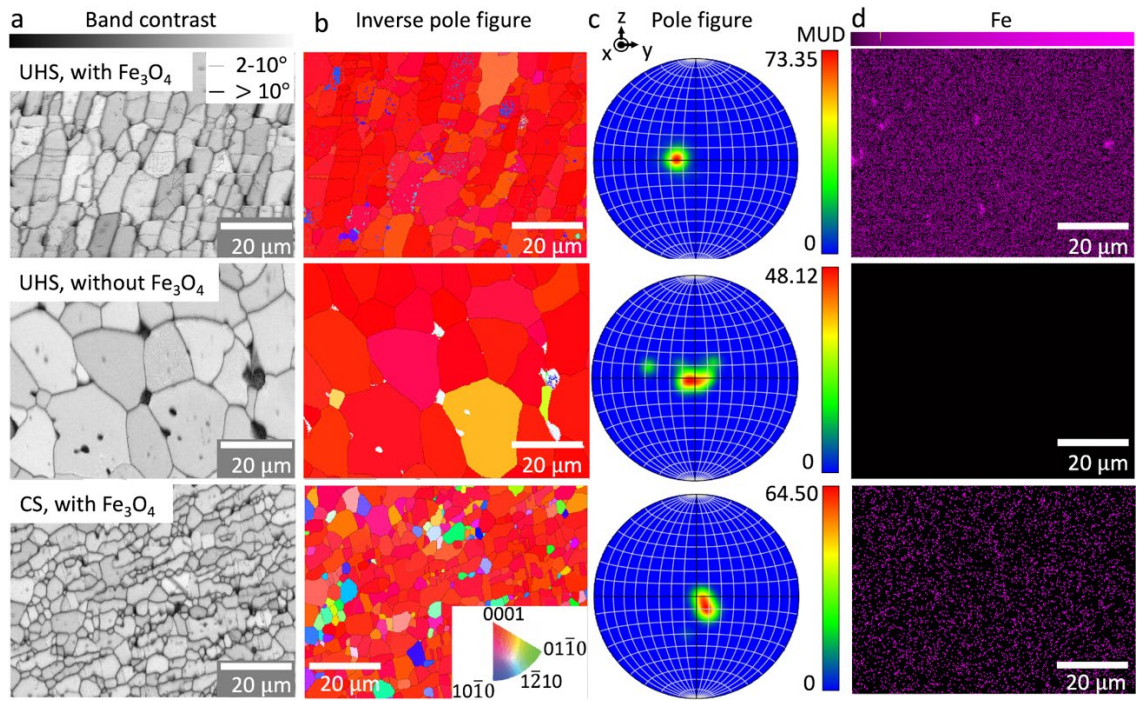

**Figure 2**: **Microstructure, texture and elemental distribution of the three types of textured alumina samples across the x-y plane (cross-section). (a)** Band contrast image showing low-angle (2-10°) and high-

angle (> 10°) grain boundaries. The brighter regions and dark regions are grain surfaces and boundaries respectively. **(b)** IPF mapping with the corresponding pole figure in **(c)** with the MUD values on the right, and **(d)** elemental mapping via EDX.

To quantify and better understand the role of the LAGB and HAGB and the Fe distributions, the microstructures and the local mechanical behaviour were studied (**Fig. 3**). The Kernel misorientation (KAM) maps obtained using EBSD at the same step-size show a high value reaching ~1° with a high concentration inside alumina grains for UHS, with $Fe_3O_4$ samples different from the other two samples that exhibited high values around grain boundaries only (**Fig. 3a, left**). This is also supported by the geometrically necessary dislocation (GND) maps with values that reach 2-3 × $10^{14}/m^2$ inside the grains of UHS, with $Fe_3O_4$ samples, which is at least $10^3$ to $10^4$ times higher than for the other samples (**Fig. 3a, right**). Such high KAM and GND values for UHS, with $Fe_3O_4$ are also observed in the y-z plane (top surface) (**Fig. S2**). The exact values from the GND maps might be unclear; however, the distribution is evident and indicative of the significant difference in GND values between samples. The GND values only indicate flow heterogeneity of crystallographic defects and are not directly indicative of plastic flow [30]. Therefore, for a detailed understanding of the role of crystallographic defects, the samples were indented at different locations on the x-y plane at least 10 times at low-load (400 mN) and high-load (1900 mN) using the load-controlled mode of the nanoindenter. It is to be noted that despite the difference in relative densities, the UHS, without $Fe_3O_4$ was stiffer, followed by 'with $Fe_3O_4$' samples for sintered by CS and UHS, indicating that $Fe_3O_4$, indeed, plays a vital role in allowing larger indentation contact depth (**Fig. 3b**). Nano-modulus ~ 230-250 GPa and Vickers hardness ~10.5-12 GPa confirm the lower values compared to the two samples (**Fig. S3**, see **Fig S4** for Vickers indents). This could be due to the lower solubility as well as insufficient time for $Fe_3O_4$ to form a solid solution with alumina in an inert sintering atmosphere [28,31].

Nevertheless, the energy dissipation measured by the area under the indentation load-contact depth curve indicates high energy dissipation for the UHS, with $Fe_3O_4$ samples, which is ~10 % and ~122 % more compared to UHS, without $Fe_3O_4$ samples, at the low and high load, respectively (**Fig. 3b,c**). The UHS, with $Fe_3O_4$ samples' energy dissipation, is also ~10 % and 99 % higher than for the CS, with $Fe_3O_4$ samples, at high and low loads, respectively (**Fig. 3b,c**). This suggests that these "crystallographic defects" might significantly contribute to the toughening of the alumina grains. However, this energy dissipation can be due to plastic events, which can be due to both plastic flow and fracture. Therefore, further nanoindentation tests were conducted on samples with $Fe_3O_4$, as both have higher energy dissipation than the pristine ones (**Fig. 3d,e**). In **Fig. 3d**, the shaded regions show the energy dissipated due to fracture events, while the rest indicates energy dissipated due to plastic flow. Indenting sequentially with increased loading from 400 mN to 1900 mN with 10 indents at each load shows that the total plastic energy dissipation increases significantly with the load for UHS, with $Fe_3O_4$ samples than for the CS, with $Fe_3O_4$ samples (**Fig. 3e**). Furthermore, the plastic energy dissipation due to cracking or voids (red, right of y-axis) nearly stays constant for both samples across different loads. This indicates that more plastic flow events occurred to dissipate energy in the UHS, with $Fe_3O_4$, likely due to the high concentration of distributed Fe atoms throughout the grains that resulted in crystallographic defects. However, there is no significant difference in the energy dissipation with loading rate within each sample containing $Fe_3O_4$ (**Fig. S5**).

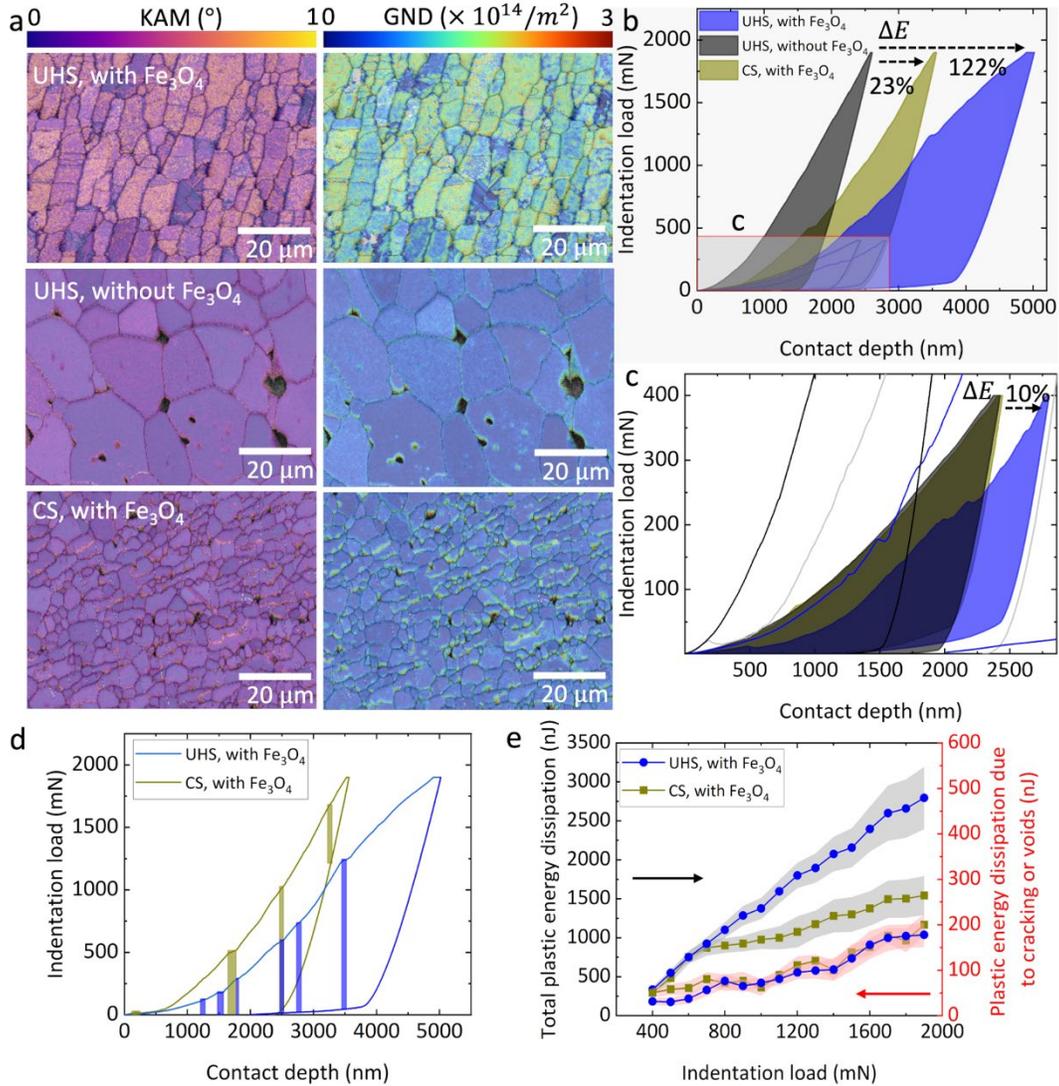

**Figure 3**: **Microstructural crystallographic defects and local mechanical behaviour of textured alumina samples.** **(a)** KAM and GND maps with respective colour codes for quantitative estimation. **(b)** Representative indentation load-contact depth plot for a low and high load of 400 mN and 1900 mN, respectively, with magnified image of low-load in **(c)**. The shaded region shows total plastic energy dissipation. **(d)** Representative indentation load-contact depth showing plastic energy dissipation due to cracking events or voids. **(e)** Measured total plastic energy dissipation and plastic energy dissipation due to cracking or voids with increasing load from 400 mN to 1900 mN. The shaded region is indicative of respective error values calculated from the area under the curve (±S.D) correlated with colour code (black: total plastic energy dissipation, red: plastic energy dissipation due to cracking events or voids).

These crystallographic defects affect the fracture toughness and flexural strength of the overall macroscopic ceramic (**Fig. 4**). The fracture toughness measured using a single-edge notched beam test setup shows rising R-curve behaviour for the UHS, with $Fe_3O_4$ samples, with $K_{Jc}$ rising from ~4.83±0.4 to ~7.0±0.3 MPa.m$^{0.5}$, till the ASTM limit (**Fig. 4a,** see **Fig. S6** for load-displacement curve, and supplementary notes for method and measurement details). In contrast, the other samples fail catastrophically, as indicated by the load-displacement curves in **Fig. S6** that have no inelastic region and thus the flat R-curves that indicate the $K_{Jc}$ is the same as crack initiation toughness ($K_{Ic}$) with low toughness values of ~4.0±0.32 MPa.m$^{0.5}$ and ~6.6±0.24 MPa.m$^{0.5}$ for UHS, without $Fe_3O_4$ and CS, with $Fe_3O_4$, respectively (**Fig. 4a**). The compliance method was used to measure the crack length from the load-displacement curves (see supplementary notes). For these two types of samples, the peak load remained constant for increasing crack lengths, which leads to $K_{Jc}=K_{Ic}$ (**Fig. S6**). In contrast, the rising R-curve behaviour in UHS, with $Fe_3O_4$ samples, can be attributed to both microstructural features, weak grain interfaces at grain boundaries and crystallographic defects within grains. Weak grain interfaces are observed from the crack deflection and multiple cracking in the electron micrographs for UHS, with $Fe_3O_4$ samples, but are absent for the others (**Fig. 4b,c**, **Fig. S7**). Furthermore, in high-contrast electron micrographs, small-scale plastic deformation at grain boundaries and slip bands emanating near the crack propagation paths within grains are revealed in the UHS, with $Fe_3O_4$ (**Fig. S7**, **Fig. 4c**, **right**). The slip bands near the crack tip are known to enhance fracture toughness in ceramics by crack tip blunting, contributing to intrinsic toughening [32]. This could also explain the enhanced $K_{Ic}$ of UHS, with $Fe_3O_4$ samples, as compared to that of UHS, without $Fe_3O_4$ [15]. Furthermore, the $K_{Ic}$ of CS, with $Fe_3O_4$ is higher than that of UHS, with $Fe_3O_4$, probably due to strong grain interfaces resulting from the long sintering time and formation of solid solutions due to increased solubility in air sintering atmosphere that could form iron aluminate spinel [28]. Lastly, the higher strength of UHS, with $Fe_3O_4$ compared to the others, could be due to the crystallographic defects or dislocation pinning at the LAGBs. This could lead to strengthening at such a meagre relative density difference of the samples [33]. Thus, the toughness vs. strength plot shows enhanced properties for the UHS, with $Fe_3O_4$ samples, as compared to the others (**Fig. 4d**). This strongly suggests that the introduction of crystallographic defects in combination with the textured microstructure and the weak interfaces that are enabled by UHS could be further enhanced intentionally to achieve even higher mechanical properties. However, more studies would be required when using different interstitial metal oxides to generalise the effect on the microstructure and resulting mechanical properties.

In summary, the effect of the $Fe_3O_4$ nanoparticles coating on alumina microplatelets for fabricating anisotropic textured alumina using MASC and UHS, named UHS, with $Fe_3O_4$ was studied. The microstructural comparison from electron diffraction shows ~25.2 % of LAGB with high texture and high Fe atom concentration compared to UHS, without $Fe_3O_4$, and CS, with $Fe_3O_4$, indicating the possible presence of crystallographic defects. Further quantitative analysis confirms this hypothesis, where higher KAM ~1° and GND ~2-3 × $10^{14}/m^2$ values are found inside alumina grains in contrast to the other samples. Energy dissipation measured from load-displacement curves using a nanoindenter indicates high energy dissipation at 400 mN and increases with increasing load to 1900 mN, reaching 122 % w.r.t UHS, without $Fe_3O_4$, and 99 % w.r.t CS, with $Fe_3O_4$. Furthermore, a comparative analysis with CS, with $Fe_3O_4$, indicates that the total plastic energy dissipated for UHS, with $Fe_3O_4$, is higher, and more energy is dissipated via plastic flow with increasing load rather than cracking. Lastly, the strength and toughness increase for UHS, with $Fe_3O_4$, due to the limited motion of these crystallographic defects at LAGBs and HAGBs, aided by extrinsic toughening due to weak grain interfaces for rising R-curve. These results indicate an effective strategy for future strong and tough ceramics by induction of 'Fe' from $Fe_3O_4$ at grain boundaries and interfaces.

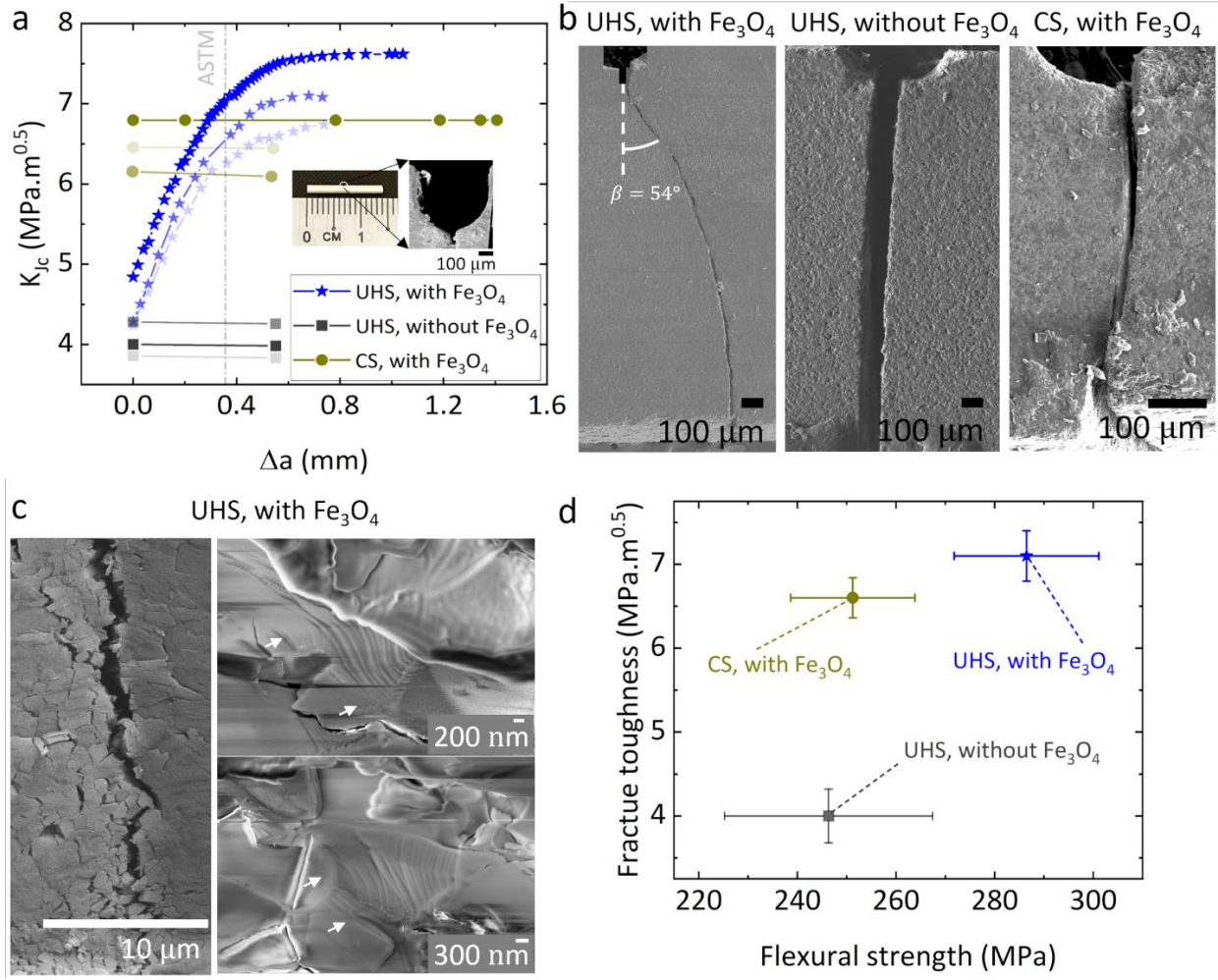

**Figure 4**: **Mechanical properties and fracture behaviour of textured alumina samples. (a)** Fracture toughness ($K_{Jc}$) as a function of the crack extension (Δa) for three test specimens in each sample category. The inset shows the prepared sample for testing. **(b)** Electron micrographs showing crack deflection and β the deflection angle in UHS, with $Fe_3O_4$. **(c)** Micro-cracking (left) and slip bands within the grain observed near the crack propagation path (right, white arrows). **(d)** Fracture toughness vs. flexural strength comparison of samples.

## Credit authorship contribution statement

**Rohit Pratyush Behera**: Conceptualization, Data curation, Formal analysis, Investigation, Methodology, Resources, Software, Validation, Visualization, Writing – original draft, Writing – review & editing. **Andrew Yun Ru Ng**: Investigation, Software, Formal analysis, Writing – review & editing. **Zehui Du**: Validation, Project administration, Resources, Writing – review & editing. **Chee Lip Gan**: Validation, Project administration, Resources, Writing – review & editing. **Hortense Le Ferrand**: Conceptualization, Formal analysis, Supervision, Funding acquisition, Project administration, Resources, Writing – review & editing.

## Declaration of competing interest

The authors declare that they have no known competing interest.

## Acknowledgements

We acknowledge the financial support from the Ministry of Education of Singapore (proposal T2EP50122-0002). We thank Prof. Ali Miserez for access to the nanoindenter and the Facility for Analysis, Characterisation, Testing and Simulation (FACTS), Nanyang Technological University, Singapore, for access to their electron microscopy facility. We acknowledge Evonik Pte Ltd for providing AERODISP® W 440.

# Supplementary Material

for

# Effect of interfacial Fe$_3$O$_4$ nanoparticles on the microstructure and mechanical properties of textured alumina densified by ultrafast high-temperature sintering

Rohit Pratyush Behera[1], Andrew Yun Ru Ng[2], Zehui Du[2,3], Chee Lip Gan[2,3], Hortense Le Ferrand[1,3*]

[1] *School of Mechanical and Aerospace Engineering, Nanyang Technological University, 50 Nanyang avenue, Singapore 639798*

[2] *Temasek Laboratories, Nanyang Technological University, 50 Nanyang Drive, Singapore 637553*

[3] *School of Materials Science and Engineering, Nanyang Technological University, 50 Nanyang avenue, Singapore 639798*

* Corresponding author: hortense@ntu.edu.sg

**Supplementary notes:**

## Experimental Section

**Sintering**

The sintering was done using two methods. Conventional sintering (CS) was carried out in a high-temperature furnace (Nabertherm, High-temperature furnace LHTCT01/16, Germany) where the samples were pre-heated at 500 °C for 5 h in the air, followed by sintering at 2.5°C/min with a dwell temperature of 1600 °C for 10 h, followed by natural cooling. Ultrafast high-temperature sintering (UHS) was conducted using a programmable power-time supply to the heating element [1]. The length of the carbon fabric in the heating element was fixed to 10 cm in length and 8 mm in breadth while keeping other parts and dimensions constant. The supplied power with time was tuned for a desirable temperature-time profile based on the conclusions from our previous work, as shown in **Figure 1b**, main manuscript [1].

**Material characterisation**

Optical images of the green bodies were taken using a digital microscope (Dino-lite AM7915MZTL, Labfriend).

The Archimedes principle was used to measure the density $\rho_S$ of the ceramics after overnight impregnation with ethanol, using:

$$\rho_S = \rho_{Eth} \times \frac{W_A}{W_{IA}-W_{IW}},$$

where $\rho_{Eth}$ is the density of ethanol at room temperature (0.789 g/cm³), $W_A$ is the dried weight of the ceramic, $W_{IA}$ is the weight of the impregnated ceramic measured in air and $W_{IW}$ is the weight of the impregnated ceramic measured in ethanol. The relative density (RD) in % was calculated using:

$$\frac{\rho_S}{\rho_{fully\ dense}} \times 100,$$

with $\rho_{fully\ dense}$ = 3.95 g/cm³. At least 5 samples were measured.

Elemental mapping (EDX) was used to characterise the distribution of the elements in the magnetised microplatelets, green body, CS samples, and UHS samples using an electron microscope (JEOL, 7800F) at 20 kV after mirror polishing.

The ceramics were mirror-polished using SiC papers (320P, 500P, 800P, 1200P, and 2400P, Struers), followed by colloidal suspensions (AP-D solution, 0.3 µm diameter Struers, and OP-U solution, 0.04 µm diameter, Struers). Crystallographic mappings were carried out using Electron Backscatter Diffraction (EBSD) on uncoated samples after polishing, using an electron microscope (JEOL, 7800F, Japan) at 20 kV and a step size of 0.21 µm. Band contrast, Kernel average misorientation (KAM), geometrically necessary dislocation (GND) maps and percentage of low angle grain boundary (LAGB) were acquired using Aztec crystal software. The bright and dark regions represent good and poor EBSD patterns, respectively. The grain sizing settings were 10 pixels per grain, with a grain detection angle of 10°, and a threshold angle of 10°.

## Local mechanical characterization

The crack paths for determining the type of fracture (intergranular or transgranular) were made using a Vickers micro-indenter (Future-Tech, FM- 300E, USA) indented on mirror polished samples at a load of 2 kgf and a dwell time of 10 s. The indents were observed under FESEM (JOEL 7600F, Japan) after gold coating. Five indents were made under the same load to observe the crack paths to ensure reliability.

Young's moduli at the nanoscale were measured using a Berkovich tip mounted nanoindenter (G200, KLA Tencor, USA) on the mirror-polished samples. Twenty indents were made with a spacing of 50 μm between each indent. The tests were performed below a thermal drift of 0.1 nm/s under load-controlled conditions. The maximum load was 200 mN, and the loading rate was kept at 1 mN/s for an indentation depth of at least 600 nm to estimate the modulus values accurately. The unloading was performed till 90 % of the loading to avoid any effects of the thermal drift on the unloading curve. The peak hold time and the Poisson's ratio were set at 10 s and 0.22, respectively. Oliver-Pharr's theory was used to obtain the Young's modulus values from the unloading curve of each indentation. The reliability of the measured values was verified using MATLAB (R2019a) to ensure that Weibull's modulus is superior to 3.

The energy dissipation was estimated using a load-indentation depth curve generated from indentation studies using a cube-corner diamond tip in a Triboindenter TI-950 nanomechanical tester (Hysitron, Bruker, Germany) for the mirror-polished samples. A high-load transducer with Hertzian contact was used for the studies performed using different maximum loads ranging from 400 mN and 1900 mN with a loading rate of 100 mN/s and a peak hold time of 10 s. All the tests were performed below a thermal drift of 0.05 nm/s.

## Bulk mechanical characterisation and calculation

**Flexural test**: The flexural properties were measured using three-point bending (3PB) test in a universal testing machine (AG-X plus, Shimadzu) with a 10 kN equipped loadcell capacity. The specimens were cut to dimensions of 15 × 1.5 × 1.5 mm$^3$ using a desktop precision diamond wire saw (STX-202A, MTI Corporation). The tests were conducted using a small 3PB jig with a load capacity of 5 kN, at a displacement rate of 1 μm.s$^{-1}$. The gauge length of the samples for the 3PB tests was 12 mm.

**Fracture toughness**: Fracture toughness tests were carried out using a single edge notch beam (SENB) setup, according to ASTM E1820-13 [2], and ASTM C1421-14 [3] standards. SENB specimens with dimensions 14 × 1.5 × 1.5 mm$^3$ was prepared using the same process as 3PB specimens first, and then the beams were notched with a 300 μm thick wire saw (STX-202A, MTI Corporation). The end of the notch was sharpened to have 20-40 μm notch diameter approximately using a razor blade coated with diamond paste (grain size ∼ 1 μm) by repeatedly passing it. Four specimens for each prepared sample having a gauge length of 12 mm were tested by monotonically loading them at a constant displacement rate of 1 μm.s$^{-1}$.

The compliance method (C) was used to determine the beginning of the crack propagation and to measure the crack length indirectly. Generally, an indirect method based on compliance evaluation is used to measure crack length during cyclic loading. However, the cyclic loading has proved unusable in this case due to small crack propagation even at low stresses upon repeated cyclic loading. Nevertheless, this approach has been used previously, where the calculated crack length ($a_n$) corresponds to the projection of the real crack length (a) starting from the position of the notch (**Equation S2**) [4–7]. The crack initiation was determined from the $f-u$ curve using (**Equation S3**) and (**Equation S4**). The crack initiation value is the point where the slope of the curve changes (deviation from the elastic region).

$$C = \frac{u}{f}, \qquad \text{(Equation S1)}$$

where $u$ is the displacement and $f$ is the force at each point after crack propagation.

The predicted crack length ($a_n$) was determined using the recursive formula.

$$a_n = a_{n-1} + \frac{W - a_{n-1}}{2} \cdot \frac{C_n - C_{n-1}}{C_n} \qquad \text{(Equation S2)}$$

where $a$ and $C$ are the crack length and the compliance calculated at the n and the $n-1$ step.

$$f\left(\frac{a}{W}\right) = \frac{3\,(a/W)^{1/2}\left[1.99 - \left(\frac{a}{W}\right)\left(1 - \left(\frac{a}{W}\right)\right)\left(2.15 - 3.93\left(\frac{a}{W}\right) + 2.7\left(\frac{a}{W}\right)^2\right)\right]}{\left[2\left(1 + \frac{2a}{W}\right)\left(1 - \frac{a}{W}\right)^{3/2}\right]} \qquad \text{(Equation S3)}$$

$$K_{Ic} = \frac{f\left(\frac{a}{W}\right) \cdot f \cdot S}{B W^{3/2}} \qquad \text{(Equation S4)}$$

where $W$ is the width, $B$ is the thickness, $S$ is the span, and $f$ is the force.

Non-linear elastic fracture mechanics analysis was used to determine the J-based crack-resistance curves ($J-R$ curves). The J corresponds to the J integral, which has both elastic and plastic contributions as described in the ASTM E1820 [2]:

$$J = J_{el} + J_{pl}, \qquad \text{(Equation S5)}$$

where $J_{el}$ and $J_{pl}$ are the elastic and plastic contributions.

The $J_{el}$ is calculated using the classic linear elastic fracture mechanics relationship:

$$J_{el} = \frac{K^2}{E'} \qquad \text{(Equation S6)}$$

where $K$ is the mode-I stress intensity factor and $E' = \frac{E}{1-\nu^2}$ for plane strain conditions.

The plane strain conditions are used in order to obtain a material-dependent stress intensity factor. The plastic component $J_{pl}$ is defined as:

$$J_{pl} = \frac{1.9\, A_{pl}}{B_N\, (W - a)} \qquad \text{(Equation S7)}$$

where $A_{pl}$ is the plastic area under the $f-u$ curve, $B_N$ is the net sample thickness, and $W-a$, is the uncracked ligament size.

The standard model $J-K$ equivalence is shown in (**Equation S8**), which is the equivalent stress intensity factor.

$$K_{JC} = \sqrt{(J_{pl} + J_{el})E'} \qquad \text{(Equation S8)}$$

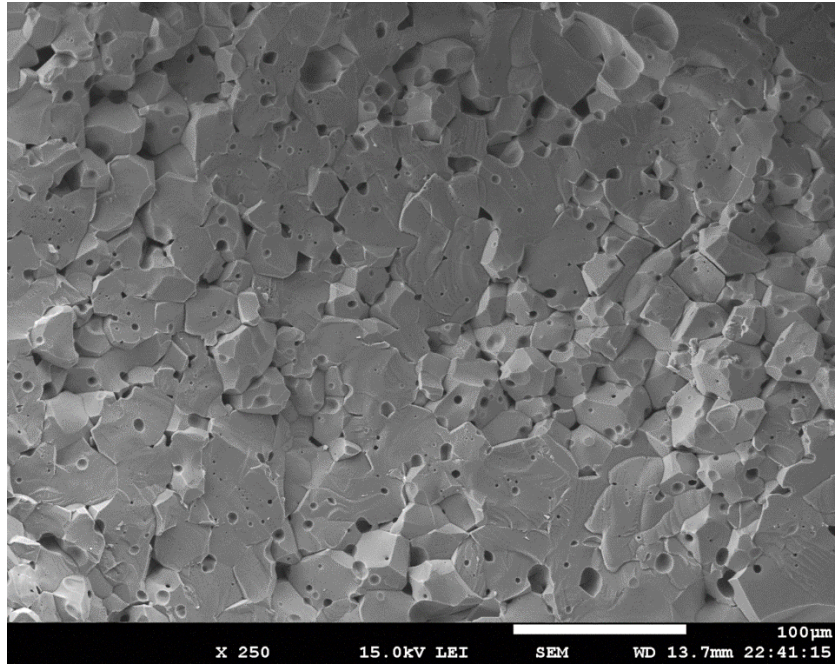

**Figure S1**: Electron micrograph for the sample UHS, without $Fe_3O_4$.

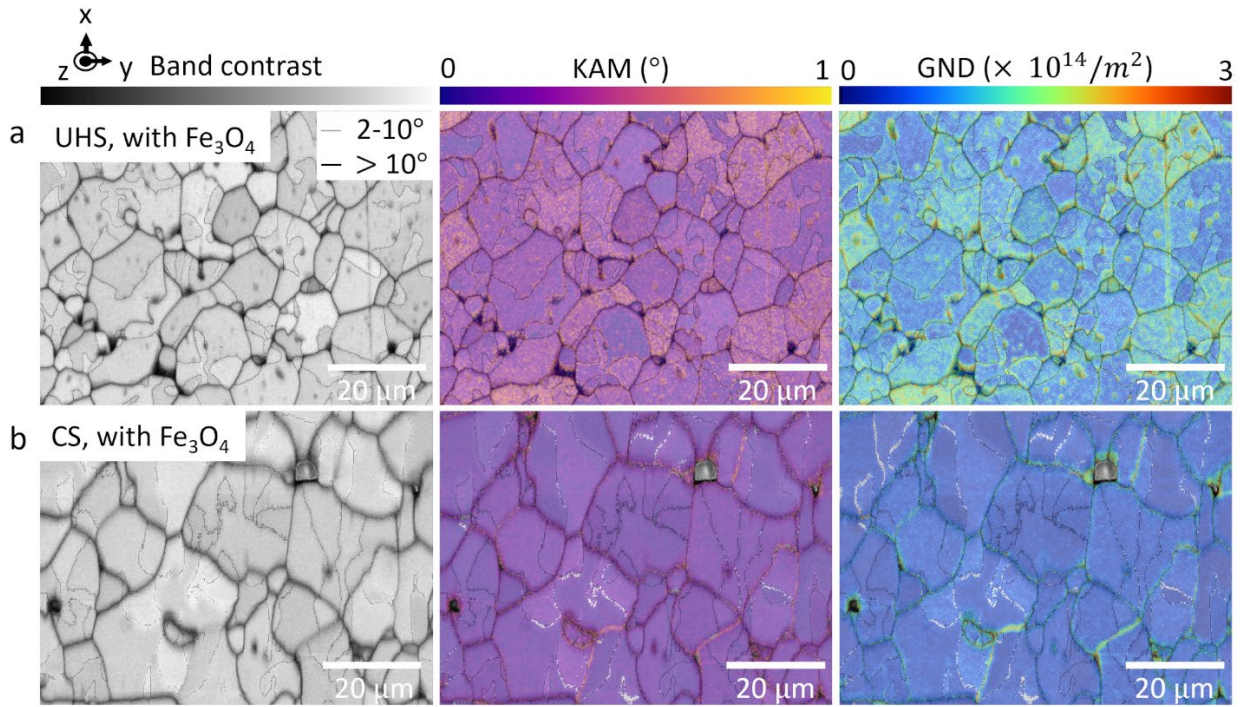

**Figure S2**: Band contrast, KAM, and GND mapping images of **(a)** UHS, with $Fe_3O_4$, and **(b)** CS, with $Fe_3O_4$, respectively. The band contrast show low-angle (2-10°) and high-angle (> 10°) grain boundaries. The brighter regions and dark regions are grain surfaces and boundaries, respectively.

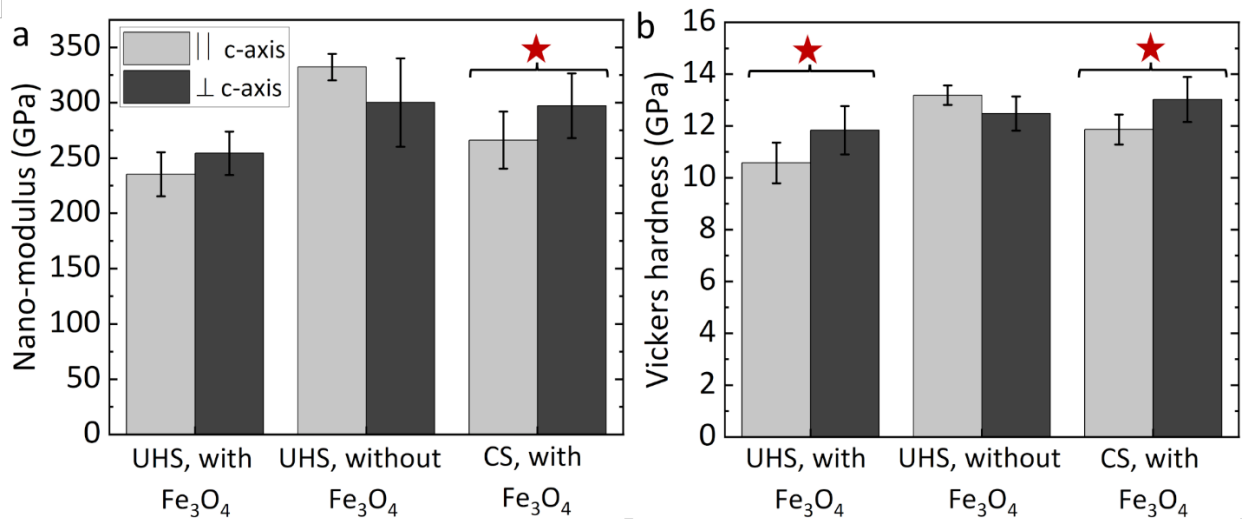

**Figure S3**: **(a)** Nano-modulus measured by nanoindentation and **(b)** Vickers hardness of the three respective samples. The symbol "★" indicates that the differences are statistically significant, with a p-value < 0.05.

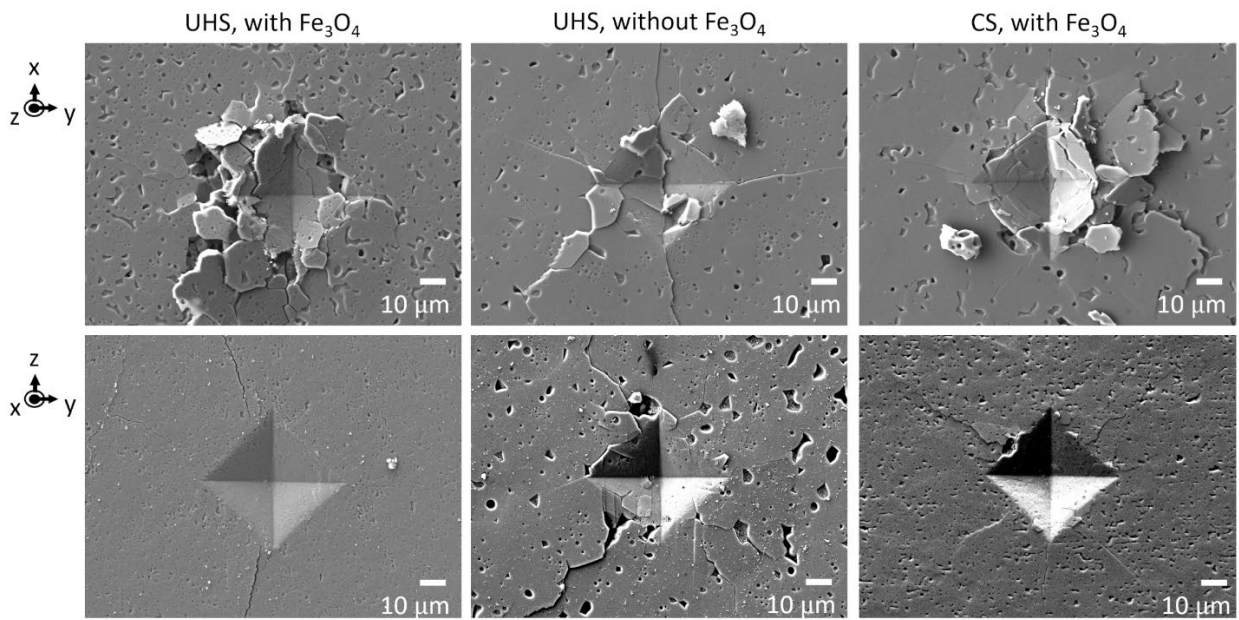

**Figure S4**: Electron micrographs of Vickers indents of the three respective samples across x-y (top) and z-y (bottom) planes.

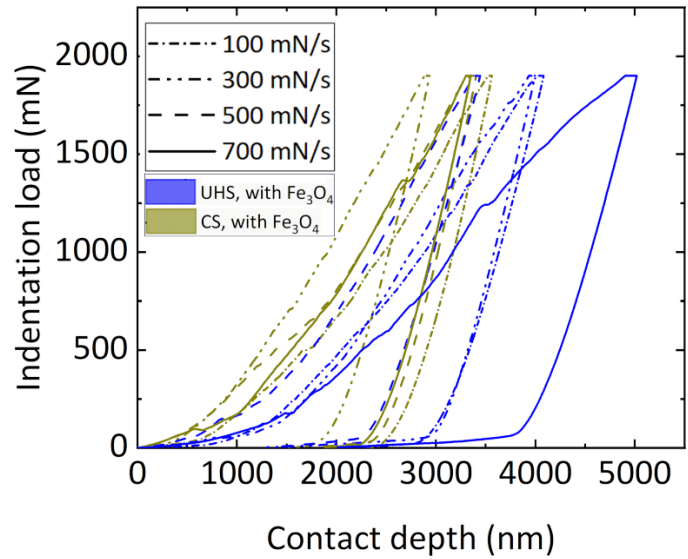

**Figure S5**: Indentation load vs. contact depth using a cube-corner tip nanoindentor for different loading rates in UHS, with $Fe_3O_4$ and CS, with $Fe_3O_4$ at a constant maximum load of 1900 mN.

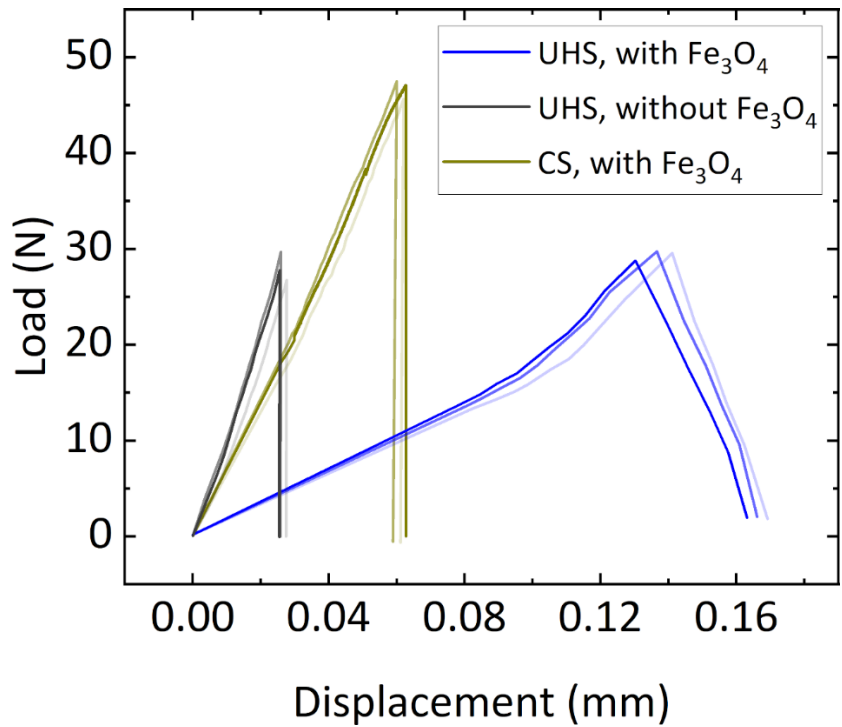

**Figure S6**: Load-displacement curves obtained from fracture toughness testing using a SENB test setup for three test specimens in each sample category.

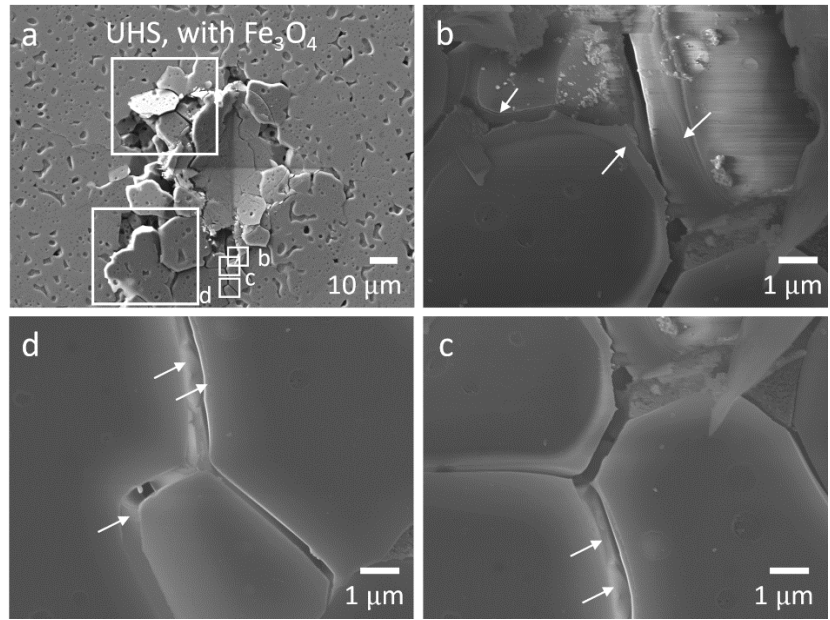

**Figure S7**: **Electron micrograph of crack path initiated by Vickers indent for UHS, with $Fe_3O_4$. (a)** Vickers indent with generated microcrack paths in **b-d**. The white box in (a) indicates pile-ups. The fracture process occurs by intergranular fracture (b-d). The white arrows in (b-d) indicate local micro-scale plastic deformations at grain boundaries.

## Supplementary References